\documentclass[acmsmall]{acmart}
\usepackage{tabularx,booktabs}
\usepackage{amsmath}
\usepackage{array}  
\usepackage{url}
\usepackage{multirow}
\usepackage{booktabs}
\usepackage{makecell} 
\usepackage{float}
\usepackage{graphicx}
\usepackage{algpseudocode} 
\usepackage[ruled,linesnumbered]{algorithm2e}
\usepackage{bm}
\AtBeginDocument{%
  }

\setcopyright{acmcopyright}
\copyrightyear{2018}
\acmYear{2018}
\acmDOI{XXXXXXX.XXXXXXX}

\acmJournal{TOIS}
\acmVolume{1}
\acmNumber{1}
\acmArticle{1}
\acmMonth{8}

\begin{document}

% \title{Defending against Adversarial Promotion in Federated Recommendation Using Purified Visual Information}
\title{Joint Semantic and Structural Representation Learning for Enhancing User Preference Modelling}

% Xuhui Ren, Wei Yuan,  Tong Chen, Chaoqun Yang, Quoc Viet Hung Nguyen, Hongzhi Yin
\author{Xuhui Ren}
\affiliation{%
  \institution{The University of Queensland}
  \city{Brisbane}
  \state{QLD}
  \country{Australia}
}
\email{xuhui.ren@uq.net.au}
\author{Wei Yuan}
\affiliation{%
  \institution{The University of Queensland}
  \city{Brisbane}
  \state{QLD}
  \country{Australia}
}
\email{w.yuan@uq.edu.au}
\author{Tong Chen}
\affiliation{%
  \institution{The University of Queensland}
  \city{Brisbane}
  \state{QLD}
  \country{Australia}
}
\email{tong.chen@uq.edu.au}
\author{Chaoqun Yang}
\affiliation{%
  \institution{Griffith University}
  \city{Gold Coast}
  \state{QLD}
  \country{Australia}
}
\email{chaoqun.yang@griffith.edu.au}
\author{Quoc Viet Hung Nguyen}
\affiliation{%
  \institution{Griffith University}
  \city{Gold Coast}
  \state{QLD}
  \country{Australia}
}
\email{henry.nguyen@griffith.edu.au}
\author{Hongzhi Yin}\authornote{Corresponding author.}
\affiliation{%
  \institution{The University of Queensland}
  \city{Brisbane}
  \state{QLD}
  \country{Australia}
}
\email{db.hongzhi@gmail.com}

\renewcommand{\shortauthors}{Ren et al.}

\begin{abstract}
  Knowledge graphs (KGs) have become important auxiliary information for helping recommender systems obtain a good understanding of user preferences. Despite recent advances in KG-based recommender systems, existing methods are prone to suboptimal performance due to the following two drawbacks: 1) current KG-based methods over-emphasize the heterogeneous structural information within a KG and overlook the underlying semantics of its connections, hindering the recommender from distilling the explicit user preferences; and 2) the inherent incompleteness of a KG (i.e., missing facts, relations and entities) will deteriorate the information extracted from KG and weaken the representation learning of recommender systems. 

  To tackle the aforementioned problems, we investigate the potential of jointly incorporating the structural and semantic information within a KG to model user preferences in finer granularity. A new framework for KG-based recommender systems, namely \textit{K}nowledge \textit{I}nfomax \textit{R}ecommender \textit{S}ystem with \textit{C}ontrastive \textit{L}earning (KIRS-CL) is proposed in this paper. Distinct from previous KG-based approaches, KIRS-CL utilizes structural and connectivity information with high-quality item embeddings learned by encoding KG triples with a pre-trained language model. These well-trained entity representations enable KIRS-CL to find the item to recommend via the preference connection between the user and the item. Additionally, to improve the generalizability of our framework, we introduce a contrastive warm-up learning strategy, making it capable of dealing with both warm- and cold-start recommendation scenarios. Extensive experiments on two real-world datasets demonstrate remarkable improvements over state-of-the-art baselines.

\end{abstract}


% \ccsdesc[300]{Security and privacy~Web application security}



\maketitle

\section{Introduction}
With the explosion of information on various online platforms (e.g., e-commerce and content-sharing platforms), recommender systems are playing an increasingly important role in alleviating the information overload for users and generating personalized recommendations to satisfy their preferences \cite{103459637482480}. A prevalent and effective paradigm for recommender systems is collaborative filtering (CF) \cite{zhu2021learning, wang2019enhancing, he2017neural}, which generates personalized recommendations based on the learned representations of users and items from historical interactions. However, they are inevitably limited by the default CF information that only involves user-item interactions, lacking the ability to model auxiliary side information like item attributes and user reviews, and underperforming in cold-start and sparse situations where users and items have few interaction records \cite{wang2019kgat}.

Among various kinds of auxiliary side information, knowledge graphs (KGs), which contain additional facts and relations about the item attributes, have gained growing attention in the revolution of traditional CF-based methods \cite{wang2020ckan, cao2019unifying, ren2021learning}. The core of a KG-based recommender is to integrate heterogeneous information into user and item representations to enhance their initial expression. Based on technical pathways to this goal, such KG-based recommenders can be categorized into four main types: feature interaction-based methods \cite{rendle2010factorization, he2017neural, liu2021contextualized}, path-based methods\cite{wang2020ckan, wang2019kgat, wang2019explainable, hu2018leveraging, wang2021learning}, regulation-based methods \cite{cao2019unifying, ai2018learning, wang2018dkn, wang2019multi} and propagation-based methods \cite{he2020lightgcn, wang2019knowledge}. However, none of them has made an effort to unify the structure and the semantic information within a KG, leading to an inevitable performance bottleneck in practice. 

\begin{figure}
	\centering
	\includegraphics[width=0.5\linewidth]{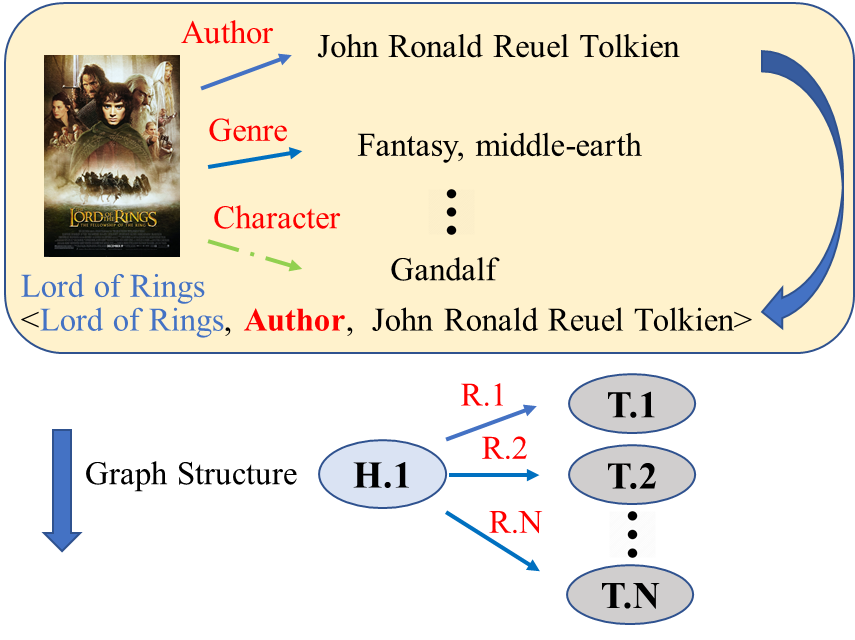}
	\caption{An illustration of the graph structure and the factual connection of KG.}~\label{fig_kg}
\end{figure}

As shown in Fig.~\ref{fig_kg}, the item entity \textit{Lord of Rings} is connected with many attribute entities (e.g., \textit{English}, \textit{Fantasy} and \textit{Gandalf}) via heterogeneous relations (e.g., \textit{author}, \textit{language} and \textit{genre}). These factual connections in KG reflect some specific aspects of the item and are presented as a subgraph structure embracing the target item entity \cite{he2020lightgcn}. To learn a comprehensive representation of the item entity, it is crucial to aggregate the information from their neighbors in the graph, where the aggregated information essentially serves as the structural representation of the target entity node. However, the learning of the structural representation fails to further explore the important semantics underlying the observed connections. Commonly, the connection information in a KG is in a head-relation-tail triple format, like \{\textit{Lord of Rings}, \textit{Author}, \textit{John Ronald Reuel Tolkien}\}. This triple connection reflects the fact that the tail entity \textit{John Ronald Reuel Tolkien} is the \textit{Author} of the head entity \textit{Lord of Rings}, where \textit{Author} is the relation between the head and tail entities. Most existing knowledge graph embedding methods usually discard the actual name of entities and relations, replacing them with unique IDs (e.g., H.x for the head entities, R.x for the relations, T.x for the tail entities) as shown in the bottom of Fig.~\ref{fig_kg} They learn the representations of entities and relations via the alignment over the graph structure. However, the name of entities and relations contain strong linguistic rules to link the common-sense knowledge, which is beneficial for the recommendation tasks. For example, given the triples \{\textit{Cake}, \textit{Ingredient}, \textit{Milk}\} and \{\textit{Chicken}, \textit{Production}, \textit{Egg}\}, the embeddings of the ''cake" and the ``egg" are dissimilar when only structural information is considered because they have no overlaps in the graph. In contrast, from semantic perspectives, \textit{Cake} and \textit{Egg} tend to have high affinity, hence deriving their semantic embeddings enables a model to explore better the hidden connections like \{Cake, Ingredient, Egg\}. Though semantics in a triple can be captured via an encoder that encodes entity and relation names \cite{kodirov2017semantic}, the encoder is prone to the biased word distribution specific to the training dataset, thus preventing the model from capturing the accurate semantic from the connection. In our paper, we point out it is necessary to leverage a language model, which is pre-trained with a large real-world corpus, to avoid learning excessive biases. And then, the connection information can be injected into this pre-trained language model for producing high-quality latent semantic representations of the KG triples \cite{yao2019kg}. With that, we aim to facilitate structural and semantic information compensating for the limited information in the user-item interaction records and making user preferences better uncovered by the recommender. 

However, the incompleteness of KGs (e.g., DBPedia\cite{auer2007dbpedia} and ConceptNet\cite{speer2017conceptnet}) limits the benefits of incorporating external knowledge into the recommendation. The movie in Fig.~\ref{fig_kg} may attract an audience due to its character \textit{Gandalf}, which is indicated by the green dashed line. However, we cannot assume any KG is complete in real applications, and such crucial connections are likely to be missed during KG construction \cite{cao2019unifying}. In this case, even though we have modeled the user preference on movies, we may still fail to recommend \textit{Lord of Rings} due to the missing connections. Therefore, it is necessary to alleviate this problem via KG completion by reasoning over potential connections. 
%As the connection in KG could be regarded as a factual description, it is promising to utilize the semantic of entities to reason their potential connections. 

In this paper, we take inspiration from the recent advances in KG embedding methods \cite{yao2019kg, liu2020k, zhang2021billion, wang2020k}, which show the potential of combining the factual connection (i.e., structural) information and the semantics of a KG to enhance entity representations. To this end, we propose a novel framework for KG-based recommendation, called Knowledge Infomax Recommender System with Contrastive Learning (KIRS-CL), which unifies the semantic and structural information of KG to improve the representation learning for recommendations. The representations facilitate the modeling of finer-grained user preferences and can discover potential KG connections in the hyperplane to guide the identification of missing links in the original KG. Specifically, KIRS-CL performs KG-based recommendations with two innovative steps. 1) Infomax KG representation learning: both the structural and semantic embeddings capture the information of KG from different perspectives and jointly contribute to higher expressiveness of the model to facilitate downstream recommendation. 2) Contrastive learning between the Infomax item representations from the KG and item embeddings from the user-item interactions (i.e., the recommendation module) to alleviate the data sparsity issue, where the users/items have few interaction records. The contrastive learning in KIRSCL enables our framework to better generalize to the cold-start scenario by preventing overfitting. 

%The structure embedding of KG is obtained via a graph convolutional network to aggregate the passing information from neighborhood entity nodes in KG.
%The heterogenous factual connection information is injected into a pretrained language model to obtain its semantic representation

To the best of our knowledge, most KG-based recommendation methods use KGs as an auxiliary attribute data source to augment the original interaction records. Furthermore, they have yet to address the aforementioned two challenges by encoding the facts with a pre-trained language model to retain a triple's semantic knowledge fully and then integrating it with graph-structured knowledge. In comparison, the main contributions of this paper are summarized as follows:
\begin{itemize}
	\item We propose a new semantic knowledge extraction method for KG-based recommendation, which integrates the observed connections (i.e., KG triples) into a language model to learn the semantic representations of entities.
	\item We devise a new framework that unifies the semantic and the structural representation of KG for capturing user preferences at a finer granularity. A contrastive warm-up learning strategy is adopted to address the cold-start problem.
	\item We conduct extensive experiments on two benchmark datasets to demonstrate the superior performance of KIRS-CL on recommendation and the KG completion tasks. 
\end{itemize}

\section{Related Work}
In this section, we present a literature review of some works related to this paper, including recommender systems, knowledge graph-based recommender systems, and contrastive learning.
% There are mainly two kinds of research push forward this work, Knowledge Graph-based recommender system and contrastive learning.
\subsection{Recommender System}
Generally, recommender systems use user-item interaction (e.g., clicks, purchases, etc.) data to model the latent features of users and items. 
The research of recommender systems can be classified into three categories: (1) matrix factorization-based methods~\cite{mnih2007probabilistic}. (2) deep learning-based methods~\cite{he2017neural}. Deep learning-based methods leverage the powerful learning ability of neural networks to model user and item features. One typical work is NCF~\cite{he2017neural}, which utilizes MLP layers to capture user and item complex relationships from massive data. (3) graph-based methods~\cite{wang2019neural,hung2017computing,xia2021self}. Graph-based methods treat each user and item as nodes and the interaction between user and item as the edge to construct a bipartite. Then, certain graph neural networks are applied to compute user and item embeddings for these nodes. For example, ~\cite{he2020lightgcn} provides a lightweight GCN model (LightGCN) to propagate information in a bipartite graph. 

Although traditional recommender systems have achieved remarkable progress, there still are some problems in these recommender systems. 
The first issue is that collaborative data, such as user-item interactions, are too sparse. As a result, there are many items in long-tail distribution. These items are cold items, which have fewer data for them to learn informative feature vectors. Therefore, the recommendation performance for these items is not good enough. 
Another problem is that existing recommender systems lack explanation. When making recommendations to a user, it will be more convincing if the system can explain why such a recommendation is made. 
To address the above issues, knowledge graphs are introduced as auxiliary information in the recommender system to provide extra information on cold items and explanations for recommendations, namely, knowledge graph-based recommender system.

\subsection{Knowledge Graph-based Recommendation}
KG recently has attracted much attention in the recommendation domain for providing rich auxiliary information about items \cite{guo2020survey}. Based on how to leverage this information for recommendations, the KG-based recommender systems can be categorized into four types, feature interaction-based methods, path-based methods, regularization-based methods and propagation-based methods. 

\begin{itemize}
\item In feature interaction-based methods, the structural knowledge in KG about the attributes of the item is directly concatenated with the item IDs to augment their feature vectors. Then, a supervised learning paradigm utilizes these knowledge-enriched item representations to calculate the recommendation score on each item. For example, \cite{piao2017factorization} extracts lightweight features from KG and inputs them into Factorization Machine \cite{rendle2010factorization}. However, the early works suffer from the problem of high dimension and sparsity of the generated feature vector. Some researchers propose to formulate a series of cross-feature interactions as combinational features to improve the model's efficiency. For example, \cite{he2017neural} analyzes the limitation of linear interactions and adopts NFM to extract high-order and non-linear feature interactions, achieving effective improvements. However, these methods can only utilize the shallow level of raw data in KG and cannot explore deeper relations among items, failing to excite the advantages of KG fully.

\item In path-based methods, multiple patterns of connections among attributes and items in KG are proposed to provide extra guidance for recommendations. Most works introduce a user-item KG to mine their relationships over the graph structure. There are mainly two directions for this approach: meta-path-based and path embedding-based methods. The meta-path-based methods utilize the connective similarities of entities in KG from different meta-paths as the regulation to constrain the representation of users and items. For example, \cite{jiang2018recommendation, wang2019kgat} propose to adopt the meta-path similarity to enrich the representation of users and items. The path embedding-based method suggests explicitly learning the embedding of paths between the user and the item in KG to model their relations. \cite{hu2018leveraging} introduces to depict the user-item interactions via the explicit representations of meta-path. However, these methods are time-consuming and labor-intensive in practice. It is not easy to find trustable rules to guide the path selection algorithm, and the effective meta-path requires domain knowledge to maintain the model fidelity.

\item In regularization-based methods, the recommender systems are regularized by additional loss from knowledge graph embedding methods for capturing the alignment relation over KG. For example, \cite{cao2019unifying} introduces TransH \cite{wang2014knowledge} to learn a translation-based model which jointly optimizes the performance of the recommendation and the KG completion in the hyperspace. \cite{wang2018dkn} includes two sources of information from the entity embedding and the word embedding to generate the news embedding for recommendation via TransD \cite{ji2015knowledge}. Although these approaches use the graph connection information to enhance the representation learning for the recommendation, their models overemphasize the connection information and ignore the underlying semantic representation of the connection fact in KG, deterring them from modeling fine-grained user preferences.

\item Propagation-based methods are based on embedding propagation with graph neural networks (GNN). These methods formulate the entity embeddings by aggregating the embedding of its multi-hop neighbors. Then the user preference can be predicted with these enriched knowledge embeddings. For example, \cite{he2020lightgcn} utilizes a convolutional neural network to enhance the representation for recommendation with its neighborhood information. \cite{wang2019knowledge} constructs the semantic embedding of the items by aggregating information from their multi-hop neighbors. 
~\cite{zou2022multi} provides three different aspects to fully use knowledge graphs, including global, local, and semantic views, achieving better performance.
~\cite{wang2023knowledge} proposes an adaptive method to avoid over-emphasis unrelated information in a knowledge graph.
~\cite{du2022hakg} attempts to apply a Hierarchy-aware graph neural network to extract underlying hierarchical information in a knowledge graph.
~\cite{xuan2023knowledge} uses a knowledge graph to deal with the sparse supervision problem.
These methods fully use the connected neighborhood entities to enrich the entity representation. However, most propagation-based methods can only study the semantic representation of entities by their neighbors, leading to an unexpected semantic bias when there are limited training instances.
\end{itemize}

We propose to tackle the aforementioned problem by extracting the semantic and the structural information to make a comprehensive modeling of the entity information. 

Besides improving recommendation performance, another advantage of incorporating a knowledge graph in the recommender system is that it can provide an explanation for a recommendation. 
~\cite{zhao2020leveraging} leverages reinforcement learning to give a demonstration for each recommendation. Specifically, it extracts imperfect path demonstrations with minimum labeling efforts and effectively leverages these
demonstrations to guide pathfinding. 
~\cite{liu2021reinforced} apply knowledge graph in news recommendation to provide knowledge-aware reasoning explanation.
~\cite{wang2022multi} utilizes both ontology-view and instance-view knowledge graphs
to model multi-level user interests; meanwhile, it leverages multi-hop path in knowledge graph to provide explanations.
~\cite{geng2022path} argues that there is ``recall bias" in the knowledge graph since not all items are reached following graph construction. Based on this argument, it provides a path language modeling recommendation framework.

\subsection{Contrastive Learning}
Due to the superior ability to regularise representations in a self-supervised manner, a resurgence of contrastive learning has been witnessed in the field of computer vision and natural language processing \cite{yu2022graph, wei2021contrastive,qiu2022contrastive,yu2022self}. An inherent advantage of contrastive learning is that it can optimize the representation and transfer the knowledge by discriminating the positive samples from the corrupted negative samples, achieving satisfactory performance when the training data is sparse \cite{wu2021self}. Some researchers propose to leverage this property to optimize the existing recommender systems in the cold start scenario. For example, \cite{wei2021contrastive} reformulates the cold-start item representation learning by maximizing the mutual dependencies between the user-item interactions and the item-feature representations. \cite{zhang2021double} tackles the data sparsity issue by proposing a double-scale dropout strategy to create contrastive signals regularising the user representation. \cite{yao2020self} introduces a two-tower DNN architecture to augment the item feature for contrastive learning. However, their modeling methods for contrastive objects are still focused on superficial attribute content or collaborative signals. We present a comprehensive entity modeling method in this work from the perspective of semantics and structure, constructing a better contrastive object to improve performance.

\section{Problem Formulation}
We first define the main concepts in our proposed KIRS-CL and then formulate our recommendation task.

\noindent
\textbf{User-Item Interaction Data}: In this paper, we mainly focus on the implicit user feedback (e.g., click, browse, purchase) on items. These feedbacks are presented as a set of interaction records $O^+=\{(u,i)\mid u\in U, i \in I\}$, where $U$ is the set of users and $I$ is the set of items. Each tuple $(u,i)\in O^+$ means user $u$ has interacted with item $i$.

\noindent
\textbf{Knowledge Graphs}: KGs contain structured side information of items (e.g., item attributes, taxonomy, and common sense knowledge). Typically, KG is composed of head-relation-tail triples of its factual connections. Each triple denotes a relation $r$ that connects the head entity $h$ to the tail entity $t$ in the graph structure, formally defined by $(h, r, t)$. A KG is the collection of these triples $G=\{(h,r,t)\mid h,t\in V, r\in R \}$, where $V$ is the set of real-world entities and $R$ is the set of relation types. With exact name matching, we can easily link the items from user interaction data and KG.

\noindent
\textbf{KG Representation Learning}: The original data of KG are stored as $G=\{(h,r,t)\mid h,t\in V, r\in R \}$. One instance $(h,r,t)$ can only express a sole connection from the head entity to the tail entity. Learning distributed representations of KG provides an effective and efficient way to preserve the nodes' connectivity in the hyperspace. We adopt TransH in this work to avoid 1-to-N, N-to-1, and N-to-N problems \cite{wang2014knowledge}. TransH assumes each relation owns a hyperplane, and a connection between the head and tail entities is valid only if both entities are projected onto the same hyperplane. It defines a distance function to indicate the possibility of a triple connection:
\begin{equation}\label{}
f(h,r,t) = \| \mathbf{h}^{\bot} +\mathbf{r}-\mathbf{t}^{\bot} \|,
\end{equation}
where $\|\cdot \|$ is the L1 distance and a lower distance indicates that the triple $(h,r,t)$ is more likely to be observed, $\mathbf{h}^{\bot}$ and $\mathbf{t}^{\bot}$ are respectively the embeddings of head and tail after projection in the hyperspace:
\begin{equation}
\begin{split}
\mathbf{h}^{\bot} = \mathbf{h}-\mathbf{w}_r^{\top}\mathbf{h}\mathbf{w}_r, \\
\mathbf{t}^{\bot} = \mathbf{t}-\mathbf{w}_r^{\top}\mathbf{t}\mathbf{w}_r,
\end{split}
\end{equation}
where $\mathbf{w}_r$ is the projection vector w.r.t. relation $r$.

\noindent
\textbf{Task Description}: Given user-item interaction records $O^+$, and KG $G$, the design target is to learn a function that predicts the probability that a user $u$ would adopt an item $i$. Based on the predicted probability, a ranking list of items can be customized for each user. 

\begin{figure*}
	\centering
	\includegraphics[width=0.9\linewidth]{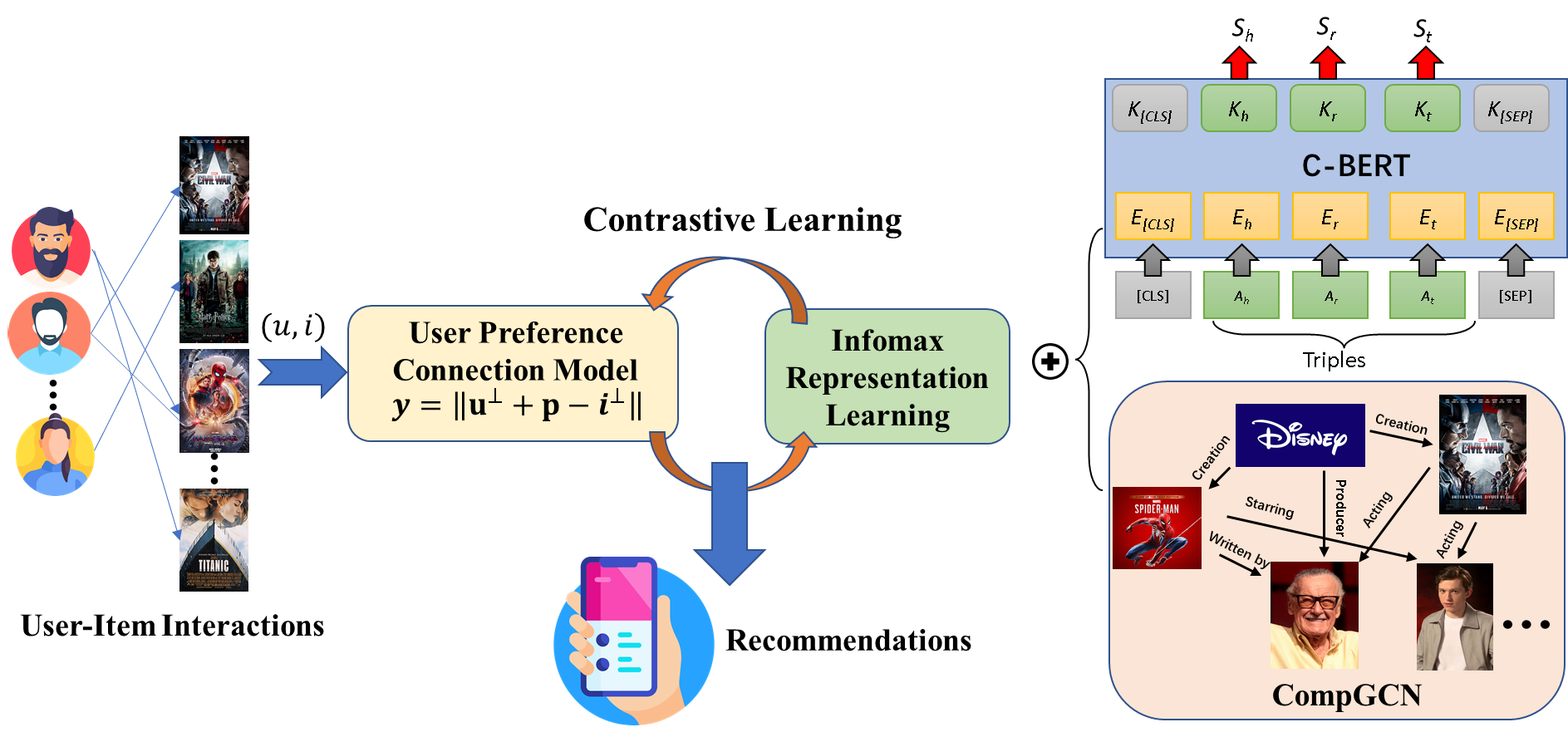}
	\caption{An illustration of our proposed KIRS-CL. The right side is the Infomax representation learning, and the left side is user preference mining. The extracted Infomax representation will combine with the user preference mining to generate the final recommendation.}
	\label{fig:diagram}
\end{figure*}

\section{Methodology}
This section describes our proposed Knowledge Infomax Recommender System with Contrastive Learning (KIRS-CL). As shown in Fig.~\ref{fig:diagram}, our proposed KIRS-CL mainly consists of two key components: 1) Infomax representation learning, which learns the semantic representation of the actual entity with a language model, and extracts the structure representation via graph convolution. 2) User preference connection model with contrastive learning, which analyses user preferences from historical user-item interactions and performs a warm-up training with contrastive learning.

\subsection{Semantic Representation Learning} \label{SRL}
KG stores structured information for items (e.g., attributes and taxonomy) in the triple format. As mentioned, a single triple in KG denotes a fact between the head and tail entities. We can regard the textual descriptions (i.e., names) of the head, relation, and tail as the contexts of each triple, and feeding them into a language model intuitively leads to a compressed representation of the semantics carried by those three components of the triple. However, simply training the language model with triples in the KG will result in language biases, as the co-occurrences of words/tokens in a KG do not necessarily reflect real-world distributions. To ensure that the language model captures the semantic meaning of each triple correctly, we propose to take advantage of pre-trained language models instead, which are trained on large-scale corpora from the web and more capable of comprehending the semantics in each triple. 

BERT \cite{devlin2018bert} has recently been widely recognized as a highly flexible, state-of-the-art pre-trained language model due to its superior performance in multiple NLP tasks. The basic structure of BERT consists of a series of multi-layer Transformer blocks \cite{vaswani2017attention}, which use the self-attention mechanism to generate the semantic embeddings of the input texts. The initial parameters of BERT are trained with 3,300M words from BooksCorpus and English Wikipedia, guaranteeing its coverage and generalization with real-world language rules in a general context. When learning the semantic representations of triples in a KG, it is highly desirable that the learned representations could maintain not only the knowledge from real-world linguistics but also the triple connections in KG. Therefore, we propose a new variant of BERT named Connection-based BERT (C-BERT) in this work to inject the connection information into the pre-trained language model.
\begin{figure*}
	\centering
	\includegraphics[width=0.6\linewidth]{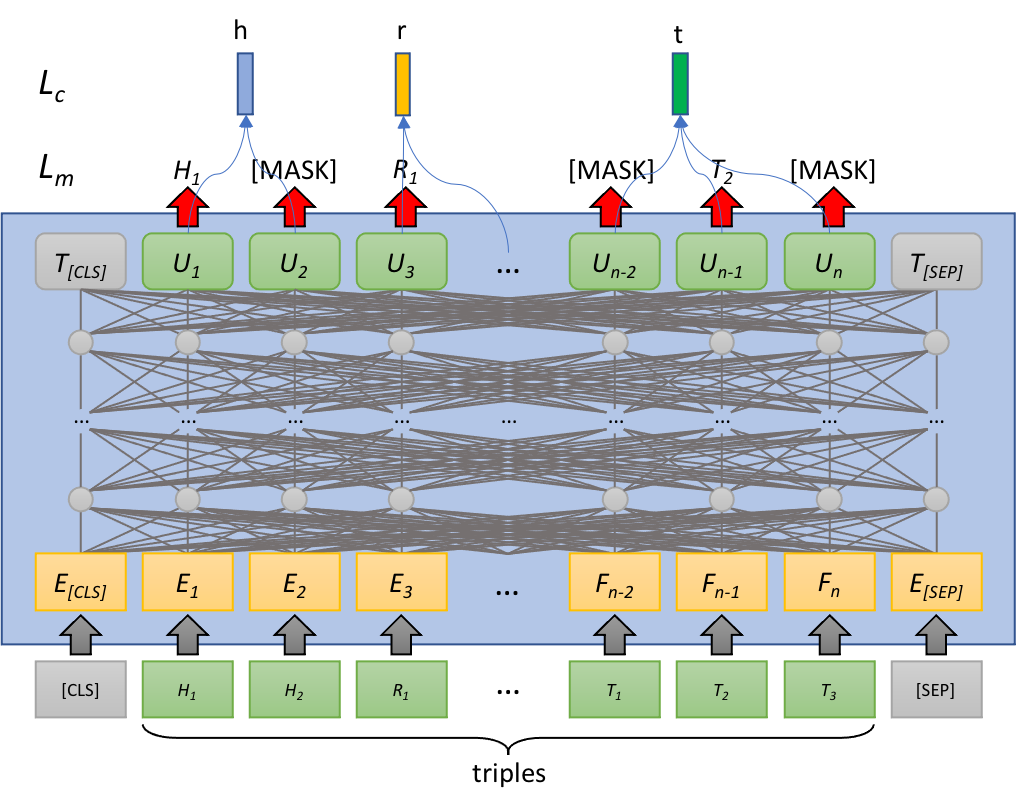}
	\caption{The structure of C-BERT. It contains two training objective: $L_{c}$ and $L_{m}$. $L_{m}$ guides C-BERT to learn semantic information from triples and $L_{c}$ enforce the learned token embeddings have spatial meaning.}
	\label{fig:diagram}
\end{figure*}

An illustration of our proposed C-BERT is shown in the upper right of Fig.~\ref{fig:diagram}. We concatenate the triple connections in KG following the order of head-relation-tail to construct a descriptive sentence and input it into BERT for fine-tuning. Following BERT's rule, the start and end tokens of the input sentences are always a classification token [CLS] and a stop sign token [SEP], respectively. As a given entity in KG may be named after several words/tokens (e.g., \textit{Lord of Rings}), we formulate the entity's input as the concatenation of its tokens, i.e., $A_x=[Tok_1^x, Tok_2^x,\dots,Tok_n^x]$. The token-level sentence expression for a triple is fed into the BERT to obtain token-wise representations. Afterward, we follow \cite{mrkvsic2017neural} to cumulate the output token embedding $\mathbf{s}$ according to the position in the input token sequence to get the corresponding representation of the head entity, the relation and the tail entity.  

We want our semantic embedding to maintain the knowledge of both the textual semantics and triple connections in KG. We design two loss functions for fine-tuning C-BERT. The first loss comes from the KG's connectivity, which requires the semantic embeddings of entity nodes to preserve the connections in the hyperspace. Thus, a margin-based loss is set to discriminate valid triples from invalid ones with the distance function in Eq.(1):
\begin{equation}\label{}
L_c=\sum_{(h,r,t)\in G}[f(\mathbf{a}_h,\mathbf{a}_r,\mathbf{a}_t)+\gamma-f(\mathbf{a}_{h},\mathbf{a}_{\hat{r}},\mathbf{a}_{\hat{t}})]_+,
\end{equation}
where $[\cdot]_+\triangleq max(0,\cdot)$, $(h, \hat{r}, \hat{t})$ is a negative sample for $(h,r,t)$ by replacing either the tail entity or the relation a random one $\hat{t}\in V$ or $\hat{r} \in R$, where $\gamma$ controls the safety margin between positive and negative triples.

The other loss for C-BERT is inspired by masked language modeling \cite{sun2019ernie}, which regards the training process as a masked token prediction task. During finetuning, each token in the input sentence has a 15\% chance of being replaced with a [MASK] token, where C-BERT is required to predict the masked tokens from inputs. It is defined as a classification task with cross-entropy loss:
\begin{equation}\label{}
L_m=-\sum_{m_i\in M}m_ilogp(\hat{m}_i),
\end{equation}
where $M$ is list of masked tokens, $p(\hat{m}_i)$ is the language model's predicted probability for token $m_i$ in the corresponding training sample. The training loss $L_g$ is the sum of $L_c$ and $L_m$ to fuse the KG connection information and the linguistic semantics into the entity representation:
\begin{equation}\label{}
L_g= L_c+L_m.
\end{equation}

\subsection{Structural Representation Learning} \label{STR}
The structured data in a KG links head and tail entities, specifically items and attributes, via relations. As shown in the bottom right of Fig.~\ref{fig:diagram}, connected entities in a KG reflect some properties of or connections to a node in the graph structure, and aggregating such information of local neighbors will help retain the structural connection and enhance the entity representation. The conventional method is to use relational graph convolutional networks (GCNs) to encode the graph-structured data with relation-specific weight matrices \cite{schlichtkrull2018modeling}. However, this approach is limited to embedding only the entities in the KG, overlooking the relation information that reflects user preferences in the recommendation scenario. Thus, in this work, we build the structural embedding module upon a composition-based multi-relational GCN (CompGCN) \cite{vashishth2019composition} to tackle the problem and obtain a comprehensive structural representation.

Given the connection data of KG $G=\{(h,r,t)\mid h,t\in V, r\in R \}$ and a convolutional network with a total of $L$ layers, we augment the original data in $G$ by swapping the head and tail entities in each triple with created reversal relations, resulting in a new KG $\hat{G}$. For each head entity $h \in V$, we can get the entity graph embedding $\mathbf{e}_h$ at the $l$-th layer ($l\leq L$) by aggregating information from $h$'s local neighbours $N_h$: 
\begin{equation}\label{}
\mathbf{e}_h^{l+1} =  {\rm ReLU}(\sum_{(t,r)\in N_h} \mathbf{W}_1^l\phi(\mathbf{e}_t^l, \mathbf{e}_r^l)+\mathbf{W}_2^l \mathbf{e}_h^l),
\end{equation}
where $\mathbf{W}_1^l$ and $\mathbf{W}_2^l$ are two learnable weight matrices at the $l$-th layer. Inspired by \cite{ji2020language}, we use a non-parametric compositional operation $\phi(\mathbf{e}_t, \mathbf{e}_r)=\mathbf{e}_t-\mathbf{e}_r$ to combine the entity and relation embeddings since they are sufficiently expressive. The relation embedding $\mathbf{e}_r$ is also transformed as:
\begin{equation}\label{}
\mathbf{e}_r^{l+1} = \mathbf{W}_3^l\mathbf{e}_r^l,
\end{equation}
where $\mathbf{W}_3^l$ is a learnable matrix at $l-th$ layer. 

We use a node prediction task to train the parameters of CompGCN with a cross-entropy loss, which distinguishes the positive triple from the negative one:
\begin{equation}\label{}
\begin{split}
L_{com}=-\sum_{(h,r,t)\in \dot{G}}ylogq([\mathbf{e}_h,\mathbf{e}_r,\mathbf{e}_t])]+\\
(1-y)log(1-q([\mathbf{e}_h,\mathbf{e}_r,\mathbf{e}_t])),
\end{split}
\end{equation}
where $q(\cdot)$ is a multi-layer perceptron (MLP) that outputs a probability of $(h,r,t)$ being true. $\dot{G}$ is the total set of positive and corrupted samples.

Finally, we concatenate the semantic representation $\mathbf{s}$ learned with C-BERT and structural representation $\mathbf{e}$ to formulate the Infomax representation of the entity in KG: $\mathbf{d} = [\mathbf{e}, \mathbf{s}]$. 

\subsection{User Preference Connection Model}
Given the user-item interaction records, we can analyze the user preferences to formulate the user embedding $\mathbf{u}$ and deduce a preference from a set of latent factors $\mathcal{P}$. All users share these factors, and each $p\in \mathcal{P}$ denotes a single aspect of preference. The extracted preference could capture the global commonality among all users that complements the local individual preference for the recommendation \cite{liu2021federated}. 

A user may adopt a recommendation due to multiple reasons. The preference $\mathbf{p}$ could be regarded as the strategy for the user when making a choice. We need to analyze which factor influences the user decision most and give more weight to the most-caring aspects for better locating the target recommendation. The score function for the user to the preference could be formulated as follows:
\begin{equation}\label{}
\begin{split}
\mathbf{p} = \sum_{p^{'}\in P} \frac{exp(log(\pi_{u,i, p^{'}}))\mathbf{p}^{'}}{\sum_{p\in P}(exp(log(\pi_{u,i, p})))},\\
\pi_{(u,i,p)} = {\rm Similarity}(\mathbf{u}+\mathbf{i}, \mathbf{p}),
\end{split}
\end{equation}

Inspired by the graph connection representation, user-item interactions could be viewed as a bipartite graph among users and items. The prediction of the recommendation for a user can be transferred to a link prediction task between the user $u$ and the item $i$ via a relation ``Preference" $p$ in the hyperplane. Similar to Eq.1, we can define the link distance from the $u$ to $i$ as follows:
\begin{equation}\label{}
j(u,i \mid p) = \| \mathbf{u}^{\bot} +\mathbf{p}-\mathbf{i}^{\bot} \|,
\end{equation}
where a lower score of $j(u,i \mid p)$ means a higher possibility of observing a link from $u$ to $i$ according to the preference $p$, which reflects $u$ will accept $i$ as the recommendation. $\mathbf{u}^{\bot}$ and $\mathbf{i}^{\bot}$ are the projection vectors in the hyperplane:
\begin{equation}
\begin{split}
\mathbf{u}^{\bot} = \mathbf{u}-\mathbf{W}_p^{\top}\mathbf{u}\mathbf{W}_p, \\
\mathbf{i}^{\bot} = \mathbf{i}-\mathbf{W}_p^{\top}\mathbf{i}\mathbf{W}_p,
\end{split}
\end{equation}
where the learned embeddings $\mathbf{i}$, $\mathbf{p}$ and $\mathbf{u}$ only depend on the interaction records, $\mathbf{W}_p$ is the projection matrix,
\begin{equation}\label{}
\mathbf{W}_p =\sum_{\hat{p} \in \mathcal{P}} \pi(u,i,\hat{p}) \mathbf{\hat{p}}.
\end{equation}

As mentioned in Section \ref{SRL} and Section \ref{STR}, we have extracted the semantic and structural information about the entities in KG, and saved them as Infomax entity representations. We can append these representations to enrich the basic information about the user and the item. Therefore, the updated representation of $i$ and $p$ could be derived as:
\begin{equation}
\begin{split}
\hat{\mathbf{i}}= \mathbf{i} + \mathbf{d}_i, \\
\hat{\mathbf{p}}= \mathbf{p} + \mathbf{d}_p,
\end{split}
\end{equation}
where $\mathbf{d}_i, \mathbf{d}_p$ are the Infomax representations of $i$ and $p$. We conduct a one-to-one mapping between the latent preference $\mathcal{P}$ and the relation $R$ to give it a practical expression. And the projection vector should be refined as:
\begin{equation}\label{}
\widehat{\mathbf{W}_p} = \mathbf{W}_p+\mathbf{W}_r,
\end{equation}

We finally adopt a BPRloss to encourage the link distance of the positive item to be smaller than the randomly selected item for each user:
\begin{equation}\label{}
L_r=\sum_{(u,i)\in O^+}\sum_{(u,i^{'}) \in O^-}-log\sigma(j(u,i^{'}\mid p^{'})-j(u,i\mid p)),
\end{equation}
where $O^{-}$ is the negative sample set generated by corrupting an interacted item to a non-interacted one.

\subsection{Contrastive Learning}
From the practical perspective, a recommender system needs to retain satisfactory performance when applied in the cold-start scenario. Therefore, we introduce contrastive learning during the training period. Generally, the cold-start problem explores how to make the recommendation when there is only a few interaction records. The lack of training samples will result in the recommender system cannot learn enough information about the user preference, thus failing to generate accurate recommendations. Some prior research proposes to utilize the attribute information in KG to augment the information of the interaction data \cite{wang2019knowledge}. However, their works lack the ability further to explore the semantic and structural knowledge of KG, failing in practical applications.

As detailed in Section \ref{SRL} and Section \ref{STR}, we have already studied comprehensive semantic and structure representations of the entity in KG. However, if we directly apply the Infomax representation to train the recommender system, the overly strong representation performance of $[\mathbf{e}_i,\mathbf{s}_i]$ will cover the original training effect of $i$ and $p$, which lead to an information loss when making the recommendation. This information loss will bring a minor influence if we have enough information to learn the user preference. But when only a few interactions are available to analyze the user preference, the user representation will highly rely on the limited training data to construct the preference connection in the hyperspace, leading to an unexpected overfitting problem.

\begin{figure}
	\centering
	\includegraphics[width=0.5\linewidth]{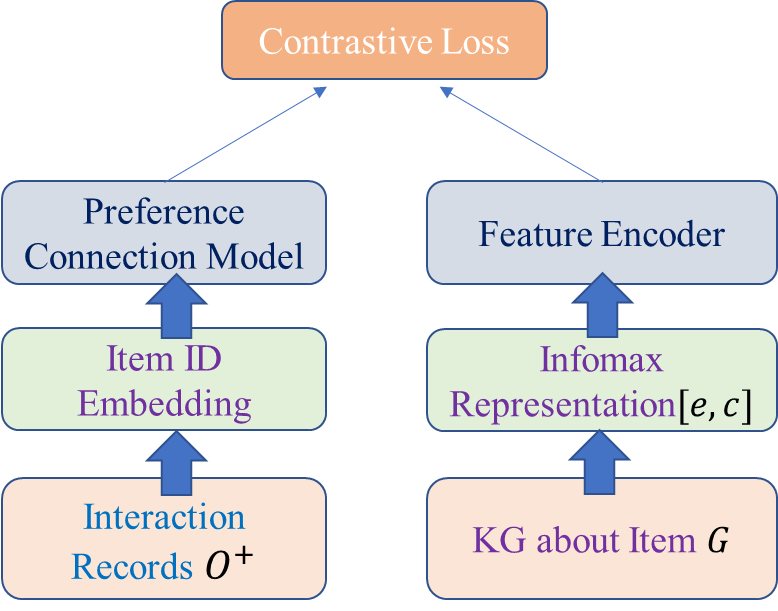}
	\caption{Illustration of contrastive learning between the item representation from the preference connection model and the Infomax representation learning.}
	\label{fig:contrastive-loss}
\end{figure}

To tackle this problem, we propose to adopt contrastive learning to warm up the initial learning process. As shown in Fig.~\ref{fig:contrastive-loss}, we set a self-discrimination task to maximize the mutual information from two different representations of items. The contrastive pairs are constructed by two sources of item representations, the item embedding $\mathbf{i}$ learnt from the preference connection model and the pretrained Infomax representation $\mathbf{d}_i$ of the item. For each contrastive pair, the item embedding $\mathbf{i}$ is set as the anchor node and is grouped with an item Infomax representation, where the paired representations refer to the same item as the positive sample and vice versa. The final contrastive set is formulated as: ${(i_1, d_1),(i_1, d_2),\cdots, (i_1, d_n)}$. The contrastive score of one instance reveals the similarity between these two sources of item representation, and the training objective is to identify the positive pairs from multiple negative pairs. As such, the learning process encourages the information transfer between two representations and provides more information bias to accelerate the learning of recommendation tasks. 

Towards representing the content information, we design a feature encoder with a two-layer MLP to process the Infomax entity representation:
\begin{equation}\label{}
\mathbf{c}_i = \mathbf{W}_5({\rm RELU}(\mathbf{W}_4\mathbf{d}_i+b_4))+b_5,
\end{equation}
where $\mathbf{W}_4$, $\mathbf{W}_5$, $b_4$, $b_5$ are learnable parameters of the encoder.

We can score the density ratio of the contrastive pairs with:
\begin{equation}\label{}
DEN_{(i,j)\mid i,j \in I}= \exp(\frac{\mathbf{i} \cdot \mathbf{c}_{j}}{\| \mathbf{i}\| \| \mathbf{c}_j\|}\cdot \frac{1}{\tau}),
\end{equation}

The final contrastive loss is formulated with InfoNCE loss \cite{poole2019variational}:
\begin{equation}\label{}
L_{con} = -\mathbb{E}_{i^{'}\in I} [log\frac{exp(\frac{\mathbf{i}^{'}\mathbf{c}_{i^{'}}}{\| \mathbf{i}^{'}\| \| \mathbf{c}_{i^{'}}\|}\cdot \frac{1}{\tau})}{\sum_{j=1}^\mathcal{J} exp(\frac{\mathbf{i}^{'}\mathbf{c}_j}{\| \mathbf{i}^{'}\| \| \mathbf{c}_j\|}\cdot \frac{1}{\tau})}],
\end{equation}
where $\mathcal{J}$ is the set of negative samples.

The final loss for warm-up training should be able to leverage contrastive learning to maximize the mutual information and conduct preference connections in the hyperspace. Therefore, the final warm-up loss is,
\begin{equation}\label{}
L_{w} = L_r+L_{con}.
\end{equation}

\subsection{Training}
The main training of our system can be separated into two parts: 1) Pretraining the Infomax representation of entities in KG, which includes the semantic representation learning and the structure representation learning; and 2) two-phase training of the recommender system, which includes a contrastive warm-up training and a recommender system training. For the first part, the training objective is to extract the semantic and structural knowledge of entities from KG. We design two losses, the entity connection loss in hyperplane $L_c$ and mask language generation loss $L_m$, to fuse the KG factual connection information and real-world language rules when generating the semantic entity representation. Then, we aggregate the information from local neighbours of the entity to learn the entity structure representation and fine-tune it with a cross-entropy loss based on a node prediction task. The well-trained entity semantic representations and structure representations are concatenated and stored as the entity Infomax representation for the downstream task. The second part starts with the contrastive warm-up learning between two sources of the item representation to enhance the learning effect with few interaction records. Then, we update the learnt representation during the warm-up period with the Infomax entity representation. At this step, the updated item representations will contain information about their contrastive pairs. The contrastive learning could be deactivated. We transfer the recommendation task as a link prediction task on the hyperplane, where a suitable recommendation should be able to link the user via the user's preference. We optimize each type of representation with BPRloss to narrow the link distance in the hyperplane. The detailed training process can be found in Algorithm 1.
\begin{algorithm}[t]
	\SetAlgoNoLine
	\KwIn{Knowledge graph $\mathcal{G}= (\mathcal{V}, \mathcal{R})$, training data $\mathcal{O}^+$, warm-up epoch $X_w$, training epoch $X$}
	Randomly initialize representation $\mathbf{u}$, $\mathbf{i}$, $\mathbf{p}$\;
	Pertrain C-BERT with Eq.(3-5) to get entity semantic representation $\mathbf{s}$; \\
	Pretrain CompGCN with Eq.(8) to get entity structure representation $\mathbf{e}$; \\
	Concatenate $\mathbf{s}$ and $\mathbf{e}$ into $\mathbf{d}$, and save locally as a dictionary $\mathcal{D}$ for entities in $\mathcal{G}$ \; 
	\For{$t\leq X_w$}{
		Warm-up training $\mathbf{u}$, $\mathbf{i}$, $\mathbf{p}$ with Eq.(10-11), Eq.(15) and Eq.(18-19) \;
	}\
	Update $\mathbf{u}$, $\mathbf{i}$, $\mathbf{p}$ with Eq.(13)\;
	\For{$X_w<t<X$}{
		Training $\mathbf{u}$, $\mathbf{i}$, $\mathbf{p}$ with Eq.(10-15)
	}\
	\KwOut{$\mathbf{u}$, $\mathbf{i}$, $\mathbf{p}$, $\mathcal{D}$}
	\caption{The training algorithm for the KIRS-CL model.}
	\label{algorithm}
\end{algorithm}

\section{Experimental Results}
We evaluate our proposed KIRS-CL on two real-world datasets, MovieLens-1m \cite{noia2016sprank} and DBook2014 \footnote{http://2014.eswc-conferences.org/important-dates/call-RecSys.html}. Specifically, we aim to answer the following research questions (RQs):
\begin{itemize}
	
\item \textbf{RQ1}: How does our proposed KIRS-CL perform compared with state-of-the-art KG-based recommender systems?

\item \textbf{RQ2}: Is our proposed Infomax representation learning really effective?

\item \textbf{RQ3}: Can KIRS-CL achieve good performance in the cold-start scenario?

\item \textbf{RQ4}: How does KIRS-CL perform in the KG completion task compared with other graph embedding methods?	
\end{itemize}

\subsection{Dataset}
For making fair experimental comparisons to demonstrate the performance of our proposed KIRS-CL,  we follow \cite{cao2019unifying} to conduct experiments on the same datasets in the movie and the book recommendation domain: MovieLens-1m and DBbook2014. Both datasets consist of users and their ratings on movies and books. We take existing ratings in datasets as the positive interaction and randomly corrupt items as the negative interaction.

KG is collected by mapping the entity in datasets into the DBpedia database via exact name matching. All triples directly related to the mapped entity are collected in our KG. The low-frequency users and items are filtered out from the datasets (i.e., the threshold is 10 for MovieLens and 5 for DBbook2014). The original DBPedia database contains many irrelevant relations to the recommendation task (e.g., wiki link and modified time). We cut off unrelated relations and filter out infrequent entities (i.e., the threshold is 10 for both datasets).

\begin{table}[]
	\caption{Statistics of MovieLens-1m and DBbook2014.}\label{tb_statistics}
	\begin{tabular}{c|c|c|c}
		\hline
		\multicolumn{2}{c|}{}                                & MovieLens & DBbook2014 \\ \hline
		\multirow{4}{*}{Interactions} & Users       & 6,040         & 5,576       \\ \cline{2-4} 
		& Items       & 3,240         & 2,680       \\ \cline{2-4} 
		& Ratings     & 998,539       & 65,961      \\ \cline{2-4} 
		& Avg.interactions
		& 165          & 12         \\ \hline
		\multirow{3}{*}{KG}        & Entity      & 14,708        & 13,882       \\ \cline{2-4} 
		& Relation    & 20           & 13          \\ \cline{2-4} 
		& Triple      & 434,189       & 334,511     \\ \hline
	\end{tabular}
\end{table}

Table~\ref{tb_statistics} shows the statistics of the processed MovieLens-1m and DBbook2014. There are 6,040 users, 3,240 items and 998,539 rating records in MovieLens-1m dataset. The average user has 165 rating records. DBbook2014 dataset has 5,576 users, 2,680 items and 6,5961 rating records, where the average user has 12 rating histories. The triple used in these two datasets is in the same format. The subgraph for Movielens-1m consists of 434,189 triples, 14,708 entities and 20 relations, while the subgraph for DBbook has 194710 triples with 8,793 entities and 9 relations.

\subsection{Baselines}
To demonstrate the effectiveness, we select six state-of-the-art baselines from KG-based recommender systems.

\begin{itemize}
\item \textbf{FM} \cite{rendle2010factorization}: This is a benchmark for the feature interaction-based models, which consider the second-order feature interactions of the input. We concatenate KG to the interaction record via exact matching of the item names, and regard it as the input.

\item \textbf{CKE} \cite{zhang2016collaborative}: This is a representative regulation-based method that adopts TransR \cite{lin2015learning} to enhance the semantic embedding in matrix factorization \cite{rendle2009bpr}.

\item \textbf{COFM} \cite{piao2018transfer}: This is a regulation-based method that regulates the training of FM with the alignment from TransE \cite{bordes2013translating}.

\item \textbf{LightGCN} \cite{he2020lightgcn}: This is a propagation-based method that utilizes a convolutional neural network to enhance the representation for recommendation with its neighborhood information.

\item \textbf{KTUP} \cite{cao2019unifying}: This is a regulation-based method that adopts TransH to formulate the translation of user preferences to recommendations.

\item \textbf{KGIN} \cite{wang2021learning}: This is a path-based method that utilizes a relation-aware information aggregation scheme to capture user preference. 
\end{itemize}

\subsection{Training Detail}
We cumulate the data in each dataset with the user ID in the interaction record. For each dataset, items for each user are randomly split with a ratio of 7:1:2 to ensure each user has at least one record in the test set.

We implement our KIRS-CL model in PyTorch. 
C-BERT is made using the PyTorch implementation of BERT made by HuggingFace~\footnote{https://github.com/huggingface/transformers}.
We use ``bert-base-uncased" as the initial point of our C-BERT model, where the embedding size is 768, and the number of transformer layers is 12.
For C-BERT, we use Adam optimizer with weight decay and learning rate of 3e-5.
The embedding size is set as 200 for CompGCN, and 968 for all other embeddings. We optimize all models (except C-BERT) with Adagrad, where the batch size is fixed as 1024. We apply a grid search for hyper-parameters: the learning rate is set: the learning rate is tuned amongst \{0.0005, 0.001, 0.005, 0.01\}, the coefficient of L2 normalization is tuned in \{$10^{-5}, 10^{-4}, 10^{-3}, 10^{-2}, 10^{-1}$\}, the temperature value $\tau$ for contrastive learning is searched in \{$0.1,0.2,0.3,\dots, 1$\}. Finally, we set the learning rate as $0.005$, the coefficient of L2 is set to $10^{-5}$, and the temperature value is set as $0.3$ for MovieLens and $0.2$ for DBbook. The number of warm-up epochs is two. Moreover, each dataset selects the first two training epochs as the warm-up training. The early stop strategy is performed (i.e., premature stopping if Precision@10 on the validation set does not increase after three training epochs). 

\begin{table}[]
\caption{Performance comparison (\%) on item recommendation ($p<0.01$) on MovieLens-1m. The best performance is shown in bold font.}\label{tb_ml_rec}
\begin{tabular}{c|ccccc}
\hline
\multirow{2}{*}{} & \multicolumn{5}{c}{MovieLens-1m}                                                                                                               \\ \cline{2-6} 
                  & \multicolumn{1}{c|}{Precision} & \multicolumn{1}{c|}{Recall}  & \multicolumn{1}{c|}{F1}      & \multicolumn{1}{c|}{Hit}     & NDCG             \\ \hline
FM                & \multicolumn{1}{c|}{29.28}     & \multicolumn{1}{c|}{11.92}   & \multicolumn{1}{c|}{13.81}   & \multicolumn{1}{c|}{81.06}   & 59.48            \\ \hline
CKE               & \multicolumn{1}{c|}{38.67}     & \multicolumn{1}{c|}{16.65}   & \multicolumn{1}{c|}{18.94}   & \multicolumn{1}{c|}{88.36}   & 67.05            \\ \hline
COFM              & \multicolumn{1}{c|}{31.74}     & \multicolumn{1}{c|}{12.74}   & \multicolumn{1}{c|}{14.87}   & \multicolumn{1}{c|}{82.67}   & 58.66            \\ \hline
LightGCN          & \multicolumn{1}{c|}{33.95}     & \multicolumn{1}{c|}{16.51}   & \multicolumn{1}{c|}{22.13}   & \multicolumn{1}{c|}{89.93}   & 65.59            \\ \hline
KTUP              & \multicolumn{1}{c|}{41.03}     & \multicolumn{1}{c|}{17.25}   & \multicolumn{1}{c|}{19.82}   & \multicolumn{1}{c|}{89.03}   & 69.92            \\ \hline
KGIN              & \multicolumn{1}{c|}{34.13}     & \multicolumn{1}{c|}{16.16}   & \multicolumn{1}{c|}{17.63}   & \multicolumn{1}{c|}{89.54}   & 65.43            \\ \hline
KIRS-CL           & \multicolumn{1}{c|}{$48.59$}   & \multicolumn{1}{c|}{$21.33$} & \multicolumn{1}{c|}{$24.54$} & \multicolumn{1}{c|}{$92.09$} & $\mathbf{77.48}$ \\ \hline
\end{tabular}
\end{table}

\begin{table}[]
\caption{Performance comparison (\%) on item recommendation ($p<0.01$) on DBbook2014. The best performance is shown in bold font.}\label{tb_dbbook_rec}
\begin{tabular}{c|ccccc}
\hline
\multirow{2}{*}{} & \multicolumn{5}{c}{DBbook2014}                                                                                                                \\ \cline{2-6} 
                  & \multicolumn{1}{c|}{Precision} & \multicolumn{1}{c|}{Recall}  & \multicolumn{1}{c|}{F1}     & \multicolumn{1}{c|}{Hit}     & NDCG             \\ \hline
FM                & \multicolumn{1}{c|}{3.44}      & \multicolumn{1}{c|}{21.55}   & \multicolumn{1}{c|}{5.75}   & \multicolumn{1}{c|}{30.15}   & 20.10            \\ \hline
CKE               & \multicolumn{1}{c|}{3.92}      & \multicolumn{1}{c|}{23.41}   & \multicolumn{1}{c|}{6.51}   & \multicolumn{1}{c|}{33.18}   & 27.78            \\ \hline
COFM              & \multicolumn{1}{c|}{3.32}      & \multicolumn{1}{c|}{20.54}   & \multicolumn{1}{c|}{5.54}   & \multicolumn{1}{c|}{28.96}   & 20.53            \\ \hline
LightGCN          & \multicolumn{1}{c|}{3.64}      & \multicolumn{1}{c|}{22.89}   & \multicolumn{1}{c|}{6.28}   & \multicolumn{1}{c|}{31.60}   & 21.55            \\ \hline
KTUP              & \multicolumn{1}{c|}{4.05}      & \multicolumn{1}{c|}{24.51}   & \multicolumn{1}{c|}{6.73}   & \multicolumn{1}{c|}{34.61}   & 27.62            \\ \hline
KGIN              & \multicolumn{1}{c|}{4.03}      & \multicolumn{1}{c|}{$25.25$} & \multicolumn{1}{c|}{6.74}   & \multicolumn{1}{c|}{34.61}   & 23.35            \\ \hline
KIRS-CL           & \multicolumn{1}{c|}{$4.58$}    & \multicolumn{1}{c|}{23.95}   & \multicolumn{1}{c|}{$7.63$} & \multicolumn{1}{c|}{$38.74$} & $\mathbf{28.19}$ \\ \hline
\end{tabular}
\end{table}

\subsection{Evaluation Metric}
The main task of this research is modeling user preference and recommending suitable items to the user. Therefore, we mainly focus on the performance of the recommendation task among different methods. The evaluation process follows \cite{cao2019unifying} that takes all items in the test sets as candidates and ranks them according to the recommendation score based on the embedding of users and items.

Five well-accepted evaluation metrics are adopted in our experiments:
\begin{itemize}
\item \textbf{Precision@K}: It evaluates the proportion of the top-K recommendation that is relevant to the user.

\item \textbf{Recall@K}: It indicates the mean probability of relevant items being successfully recommended to the user.

\item \textbf{F1 score@K}: It is a combinational evaluation of precision@K and recall@K.

\item \textbf{Hit@K}: It shows whether the relevant item is recommended within the top-K recommendation list.

\item \textbf{nDCG@K}: Normalized Discounted Cumulative Gain is a standard metric to measure the order of the positive and negative results in the top-K recommendation list. 
\end{itemize}
We set $K=10$ in the experiment to examine the recommendation performance on top-10 results.

\subsection{RQ1: Performance Comparision}
Table~\ref{tb_ml_rec} and \ref{tb_dbbook_rec} report the overall performance of our proposed KIRS-CL and baselines on two datasets. We can observe that our proposed model achieves a surprising performance gain on every evaluation metric compared with other GNN-based and KG-based models. This is mainly because our proposed Infomax entity representation has fully explored the semantic and structural information of KG, which strongly supports our system to make a good understanding of user preference. The Infomax representation incorporates real-world linguistic rules and KG connections to enhance the initial representation of entities in KG, making it easier for our system to find target recommendations via preference connections in the hyperplane. Also, from the performance comparison between MovieLens and DBbook, where the average number of interactions for each user is 165 in MovieLens and 12 in DBbook, our proposed method obtains more performance gains and shows much better recommendation performance when there are sufficient interaction data. 

Most regulation-based and path-based methods (i.e., CKE, KTUP and KGIN) perform better than the typical feature interaction-based method (i.e., FM). That is mainly because the regulation-based and the path-based method can learn a better entity representation with the graph structure information beyond the contextual attribute information. However, the regulated-based COFM performs worse than FM on the DBbook dataset. One possible reason is that the inherent drawback of TransE (i.e., 1-to-N and N-to-N issues \cite{wang2014knowledge}) cannot support a good representation learning when there are limited training interactions. 

KTUP achieves robust and satisfying performance on both datasets, mainly because it designs a joint training of the triple connection with TransH and the user representation learning. This joint-training mode can conduct the recommendation prediction and the KG completion seamlessly, leading to better recommendation results.

All models perform much better on MovieLens than on DBbook due to the relatively sufficient training data.
\begin{table}[]
\caption{The recommendation performance comparison (\%) among variants of KIRS on MovieLens-1m.}\label{tb_ml_ablation}
\begin{tabular}{c|ccccc}
\hline
\multirow{2}{*}{} & \multicolumn{5}{c}{MovieLens-1m}                                                                                                 \\ \cline{2-6} 
                  & \multicolumn{1}{c|}{Precision} & \multicolumn{1}{c|}{Recall} & \multicolumn{1}{c|}{F1}    & \multicolumn{1}{c|}{Hit}     & NDCG  \\ \hline
KIRS-CL           & \multicolumn{1}{c|}{48.59}     & \multicolumn{1}{c|}{21.33}  & \multicolumn{1}{c|}{24.54} & \multicolumn{1}{c|}{$92.09$} & 77.48 \\ \hline
KIRS/se           & \multicolumn{1}{c|}{45.32}     & \multicolumn{1}{c|}{19.43}  & \multicolumn{1}{c|}{22.44} & \multicolumn{1}{c|}{90.45}   & 74.18 \\ \hline
KIRS/st           & \multicolumn{1}{c|}{47.17}     & \multicolumn{1}{c|}{20.56}  & \multicolumn{1}{c|}{23.64} & \multicolumn{1}{c|}{91.37}   & 76.37 \\ \hline
KIRS/fi           & \multicolumn{1}{c|}{45.49}     & \multicolumn{1}{c|}{19.58}  & \multicolumn{1}{c|}{22.59} & \multicolumn{1}{c|}{90.72}   & 74.89 \\ \hline
KIRS/In           & \multicolumn{1}{c|}{37.00}     & \multicolumn{1}{c|}{16.79}  & \multicolumn{1}{c|}{18.76} & \multicolumn{1}{c|}{89.47}   & 67.02 \\ \hline
\end{tabular}
\end{table}

\begin{table}[]
\caption{The recommendation performance comparison (\%) among variants of KIRS on DBbook2014.}\label{tb_dbbook_ablation}
\begin{tabular}{c|ccccc}
\hline
\multirow{2}{*}{} & \multicolumn{5}{c}{DBbook2014}                                                                                                                \\ \cline{2-6} 
                  & \multicolumn{1}{c|}{Precision} & \multicolumn{1}{c|}{Recall}  & \multicolumn{1}{c|}{F1}     & \multicolumn{1}{c|}{Hit}     & NDCG             \\ \hline
KIRS-CL           & \multicolumn{1}{c|}{$4.58$}    & \multicolumn{1}{c|}{$23.95$} & \multicolumn{1}{c|}{$7.63$} & \multicolumn{1}{c|}{$38.74$} & $\mathbf{28.19}$ \\ \hline
KIRS/se           & \multicolumn{1}{c|}{4.13}      & \multicolumn{1}{c|}{23.51}   & \multicolumn{1}{c|}{6.87}   & \multicolumn{1}{c|}{35.15}   & 25.91            \\ \hline
KIRS/st           & \multicolumn{1}{c|}{4.46}      & \multicolumn{1}{c|}{23.31}   & \multicolumn{1}{c|}{7.39}   & \multicolumn{1}{c|}{37.53}   & 26.69            \\ \hline
KIRS/fi           & \multicolumn{1}{c|}{4.07}      & \multicolumn{1}{c|}{22.83}   & \multicolumn{1}{c|}{7.05}   & \multicolumn{1}{c|}{35.89}   & 26.11            \\ \hline
KIRS/In           & \multicolumn{1}{c|}{3.62}      & \multicolumn{1}{c|}{22.81}   & \multicolumn{1}{c|}{6.06}   & \multicolumn{1}{c|}{31.42}   & 21.54            \\ \hline
\end{tabular}
\end{table}
\subsection{RQ2: Ablation Study}
We further conduct some ablation studies to verify the actual effectiveness of our proposed modules. The main contributions of this work are the construction of the Infomax representation and the contrastive learning strategy. In this experiment, we first create a plain variant of the default setting \textbf{KIRS-CL} that does not include contrastive learning and Infomax representation as \textbf{KIRS/In}. We report the performance improvement of each variant over the plain version.

KIRS-CL introduces a powerful Infomax representation. The Infomax representation includes semantic representation learning and structure representation learning. Conditioned on the plain variant KIRS/In, we separately add the semantic and structural representations to validate each technical component's effect. We name the plain version plus semantic representation as \textbf{KIRS/st} and the plain version plus structure representation as \textbf{KIRS/se}. In addition, to demonstrate the improvement of our fine-tuning strategy on C-BERT, we design a variant to extract the semantic information only based on BERT, denoted as \textbf{KIRS/fi}.

From Table~\ref{tb_ml_ablation} and~\ref{tb_dbbook_ablation}, we can witness that the default setting of KIRS-CL achieves the best performance among all variants, that is mainly because KIRS-CL can make full use of the semantic the structure information of KG to enhance the user preference modeling. Both these two perspectives of information can provide auxiliary information from different views, which enables our system better distill the information in KG.

Comparing KIRS/se and KIRS/st to the plain variant KIRS/In, we can see our extracted semantic representation and the structure representation effectively improve the recommendation performance. The attachment of semantic representation and the structure representation achieves remarkable recommendation performance gains in all evaluation metrics. That is mainly because each representation stands for a unique perspective to distill the information of KG for recommendation predictions. In the meantime, we notice that the variant with semantic representation KIRS/st performs better than the variant with structure representation. One possible reason is that the base module to generate the semantic representation is BERT, trained with tremendous web resources. It performs better in modeling the context semantics with real-world linguistic rules. Our proposed C-BERT takes the factual triple connection in KG as the input sentence and gets fine-tuned with $L_m$ and $L_c$ (ref. Section \ref{SRL}). The improvements of KIRS/st based on KIRS/fi further verify that our fine-tuning of C-BERT could achieve a better information extraction performance than the base BERT model. The final generated semantic representation has established the semantic connection relation in the hyperspace, making it more suitable for our recommendation task.

\subsection{RQ3: Cold-start Experiments}
We further verify the performance of our warm-up training strategy in the cold-start application scenario. The DBbook dataset is relatively small, and there is no more difference if we split this dataset. Therefore, the cold-start experiment is mainly focusing on the MovieLens dataset. We create a new variant that excludes the contrastive warm-up learning process, denoted as \textbf{KIRS}. We cumulate interactions in the training set of MovieLens with the user ID and randomly split the item for each user into four subsets. We concatenate the subset according to different experimental targets to conduct the training with 25\%, 50\%, and 75\% of the training samples.
\begin{table}[]
	\centering
	\caption{The recommendation performance (\%) with 25\% training samples.}\label{tb_25}
	\begin{tabular}{c|ccccc}
		\hline
		\multirow{2}{*}{} & \multicolumn{5}{c}{MovieLens-1m, 25\% interactions}                                                                            \\ \cline{2-6} 
				& \multicolumn{1}{c|}{Precision} & \multicolumn{1}{c|}{Recall} & \multicolumn{1}{c|}{F1} & \multicolumn{1}{c|}{Hit} &NDCG  \\ \hline
		FM		& \multicolumn{1}{c|}{15.73} & \multicolumn{1}{c|}{7.64} & \multicolumn{1}{c|}{8.26} & \multicolumn{1}{c|}{72.75} &43.79  \\ \hline				
		CKE		& \multicolumn{1}{c|}{12.45} & \multicolumn{1}{c|}{6.25} & \multicolumn{1}{c|}{6.75} & \multicolumn{1}{c|}{65.50} &37.85  \\ \hline
		COFM		& \multicolumn{1}{c|}{15.79} & \multicolumn{1}{c|}{7.94} & \multicolumn{1}{c|}{8.57} & \multicolumn{1}{c|}{73.39} &44.08  \\ \hline
		LightGCN		& \multicolumn{1}{c|}{15.23} & \multicolumn{1}{c|}{8.72} & \multicolumn{1}{c|}{11.04} & \multicolumn{1}{c|}{75.52} &43.63  \\ \hline
		KTUP		& \multicolumn{1}{c|}{19.50} & \multicolumn{1}{c|}{6.15} & \multicolumn{1}{c|}{7.59} & \multicolumn{1}{c|}{63.71} &45.03  \\ \hline
		KGIN		& \multicolumn{1}{c|}{15.07} & \multicolumn{1}{c|}{8.51} & \multicolumn{1}{c|}{8.90} & \multicolumn{1}{c|}{75.96} &44.22  \\ \hline
		KIRS		& \multicolumn{1}{c|}{23.52} & \multicolumn{1}{c|}{9.36} & \multicolumn{1}{c|}{10.82} & \multicolumn{1}{c|}{77.14} &$\mathbf{54.12}$  \\ \hline
		KIRS-CL		& \multicolumn{1}{c|}{$\mathbf{26.42}$} & \multicolumn{1}{c|}{$\mathbf{10.26}$} & \multicolumn{1}{c|}{$\mathbf{12.03}$} & \multicolumn{1}{c|}{$\mathbf{77.93}$} &53.92  \\ \hline
	\end{tabular}
\end{table}

Table~\ref{tb_25} presents the recommendation performance of baselines and our proposed method with 25\% of the training set. We can observe that our proposed warm-up contrastive learning is effective when there are limited interaction records and help our model achieve the best performance among all methods. All methods suffer from a performance loss when decreasing the instance of training samples. Although KIRS does not involve contrastive warm-up learning, our Infomax representation can still offer enough information to model user preference, achieving the second-best of all methods. 
\begin{table}[]
	\centering
	\caption{The recommendation performance with 50\% training samples.}\label{tb_50}
	\begin{tabular}{c|ccccc}
		\hline
		\multirow{2}{*}{} & \multicolumn{5}{c}{MovieLens-1m, 50\% interactions}                                                                            \\ \cline{2-6} 
		& \multicolumn{1}{c|}{Precision} & \multicolumn{1}{c|}{Recall} & \multicolumn{1}{c|}{F1} & \multicolumn{1}{c|}{Hit} &NDCG  \\ \hline
		FM		& \multicolumn{1}{c|}{18.91} & \multicolumn{1}{c|}{9.08} & \multicolumn{1}{c|}{10.00} & \multicolumn{1}{c|}{72.75} &43.79  \\ \hline				
		CKE		& \multicolumn{1}{c|}{20.84} & \multicolumn{1}{c|}{9.95} & \multicolumn{1}{c|}{9.95} & \multicolumn{1}{c|}{75.93} &48.28  \\ \hline
		COFM		& \multicolumn{1}{c|}{19.79} & \multicolumn{1}{c|}{10.13} & \multicolumn{1}{c|}{10.84} & \multicolumn{1}{c|}{79.83} &49.44  \\ \hline
		LightGCN		& \multicolumn{1}{c|}{19.11} & \multicolumn{1}{c|}{10.90} & \multicolumn{1}{c|}{13.83} & \multicolumn{1}{c|}{82.28} & 50.09 \\ \hline
		KTUP		& \multicolumn{1}{c|}{23.75} & \multicolumn{1}{c|}{9.02} & \multicolumn{1}{c|}{10.51} & \multicolumn{1}{c|}{75.61} &52.14  \\ \hline
		KGIN		& \multicolumn{1}{c|}{19.06} & \multicolumn{1}{c|}{10.81} & \multicolumn{1}{c|}{11.22} & \multicolumn{1}{c|}{82.40} &49.81  \\ \hline
		KIRS		& \multicolumn{1}{c|}{32.77} & \multicolumn{1}{c|}{13.35} & \multicolumn{1}{c|}{15.46} & \multicolumn{1}{c|}{84.06} & 63.80 \\ \hline
		KIRS-CL		& \multicolumn{1}{c|}{$\mathbf{33.23}$} & \multicolumn{1}{c|}{$\mathbf{13.60}$} & \multicolumn{1}{c|}{$\mathbf{15.72}$} & \multicolumn{1}{c|}{$\mathbf{84.95}$} &$\mathbf{64.65}$  \\ \hline
	\end{tabular}
\end{table}

\begin{table}[]
		\centering
	\caption{The recommendation performance (\%) with 75\% training samples.}\label{tb_75}
	\begin{tabular}{c|ccccc}
		\hline
		\multirow{2}{*}{} & \multicolumn{5}{c}{MovieLens-1m, 75\% interactions}                                                                            \\ \cline{2-6} 
		& \multicolumn{1}{c|}{Precision} & \multicolumn{1}{c|}{Recall} & \multicolumn{1}{c|}{F1} & \multicolumn{1}{c|}{Hit} &NDCG  \\ \hline
		FM		& \multicolumn{1}{c|}{23.96} & \multicolumn{1}{c|}{10.59} & \multicolumn{1}{c|}{12.01} & \multicolumn{1}{c|}{79.98} & 53.63 \\ \hline				
		CKE		& \multicolumn{1}{c|}{25.46} & \multicolumn{1}{c|}{11.36} & \multicolumn{1}{c|}{12.58} & \multicolumn{1}{c|}{81.87} &54.27  \\ \hline
		COFM		& \multicolumn{1}{c|}{24.28} & \multicolumn{1}{c|}{11.69} & \multicolumn{1}{c|}{12.81} & \multicolumn{1}{c|}{82.78} &54.81  \\ \hline
		LightGCN		& \multicolumn{1}{c|}{24.28} & \multicolumn{1}{c|}{13.05} & \multicolumn{1}{c|}{16.92} & \multicolumn{1}{c|}{85.98} &55.18  \\ \hline
		KTUP		& \multicolumn{1}{c|}{25.46} & \multicolumn{1}{c|}{11.36} & \multicolumn{1}{c|}{12.58} & \multicolumn{1}{c|}{81.87} &54.27  \\ \hline
		KGIN		& \multicolumn{1}{c|}{24.49} & \multicolumn{1}{c|}{12.94} & \multicolumn{1}{c|}{13.71} & \multicolumn{1}{c|}{86.29} &55.34  \\ \hline
		KIRS		& \multicolumn{1}{c|}{43.54} & \multicolumn{1}{c|}{$\mathbf{18.64}$} & \multicolumn{1}{c|}{21.45} & \multicolumn{1}{c|}{$\mathbf{90.03}$} &72.77  \\ \hline
		KIRS-CL		& \multicolumn{1}{c|}{$\mathbf{43.90}$} & \multicolumn{1}{c|}{18.81} & \multicolumn{1}{c|}{$\mathbf{21.64}$} & \multicolumn{1}{c|}{89.95} &$\mathbf{73.02}$  \\ \hline
	\end{tabular}
\end{table}

The recommendation performance from Table~\ref{tb_25} to Table~\ref{tb_75} indicates the importance of the training sample to learn user preference. The sufficient training data will help the recommender system better understands user preferences and generates suitable recommendations. In these three batches of the experiment, our proposed KIRS-CL and KIRS consistently perform the best two results. The remarkable performance gain in cold-start scenarios effectively demonstrates our proposed system can explore user preference from limited interaction records with Infomax representation and generate suitable recommendations with the preference connection model. The comparison between KIRS and KIRS-CL on 25\% and 50\% training experiments indicates that contrastive warm-up learning can improve performance when the training samples are limited. But the influence of contrastive warm-up training will recede with the increment of the training samples.

An interesting phenomenon is the method that utilizes the information propagation on the recommendation path or information from their neighbors (i.e., LightGCN and KGIN) could also maintain a good recommendation performance. That is mainly because these methods do not require forcing their embeddings to be smaller with the aligned items in the hyperplane. Therefore, they can avoid the learning error from the alignment task when there are few interactions.
 
\subsection{RQ4: KG Completion}
In this section, we want to examine whether our proposed KIRS-CL has the ability to revise the missing entities/links over KG, and tackle the incompleteness of KG. We design a KG completion task that predicts the missing head entity $h$ or the tail entity $t$ for a given triple $(h,r,t)$. All entities in KG is regarded as a candidate to fill the missing position in the triple. We rank the candidates according to the score calculated with entity embeddings and relation embeddings \cite{bordes2014semantic}.  

For convenience, we further introduce two widely accepted evaluation metrics for link prediction \cite{yao2019kg}:
\begin{itemize}
\item \textbf{Hit@K}: It shares a similar definition as in the recommendation task. The top-K entity prediction that contains the missing entity in the triple could be regarded as a positive prediction. We compute the result of all triple predictions as the hit ratio score.

\item \textbf{Mean Rank}: It calculates the mean rank order of the missing entities. A lower value means the correct prediction has a better ranking performance.
\end{itemize}

The KG-based methods, which can model the graph connection information, are selected as baselines in this section. We also add the comparison with benchmark methods, for example, TransE, TransH, and TransR.

\begin{table}[]
		\centering
	\caption{Performance comparison on KG completion task.}\label{tb_completion}
	\begin{tabular}{c|cc|cc}
		\hline
		\multirow{2}{*}{} & \multicolumn{2}{c|}{MovieLens-1m} & \multicolumn{2}{c}{DBbook2014}  \\ \cline{2-5} 
		& \multicolumn{1}{c|}{Hit@10}  & MR & \multicolumn{1}{c|}{Hit@10} & MR \\ \hline
		TransE            & \multicolumn{1}{c|}{46.95}        &537    & \multicolumn{1}{c|}{60.71}       &531    \\ \hline
		TransH            & \multicolumn{1}{c|}{47.63}        &537    & \multicolumn{1}{c|}{60.06}       &556    \\ \hline
		TransR            & \multicolumn{1}{c|}{38.93}        & 609   & \multicolumn{1}{c|}{56.33}       &563    \\ \hline
		CKE               & \multicolumn{1}{c|}{34.37}        &585    & \multicolumn{1}{c|}{54.66}       &593    \\ \hline
		CoFM              & \multicolumn{1}{c|}{46.51}        &506    & \multicolumn{1}{c|}{60.81}       &521    \\ \hline
		KTUP              & \multicolumn{1}{c|}{$\mathbf{48.90}$}        & 527   & \multicolumn{1}{c|}{60.75}       &499    \\ \hline
		KIRS-CL           & \multicolumn{1}{c|}{44.39}        &$\mathbf{160}$    & \multicolumn{1}{c|}{$\mathbf{61.58}$}       &$\mathbf{244}$    \\ \hline
	\end{tabular}
\end{table}

Table~\ref{tb_completion} presents the KG completion performance of all methods. It is obvious that our proposed KIRS outperforms all other baselines on both datasets except the hit@K value on MovieLens. That is mainly because our proposed C-BERT could enhance the structure connection via real-world linguistic rules. Different from the previous alignment-based methods, KIRS-CL could fuse the real-world semantic information and the graph connection information to enhance the basic information expression of the entity node in KG. The extracted semantic information could help our system to construct a latent connection in the hyperplane to link the entities with specific relations. We argue the weak performance of Hit@K on MovieLens is primarily because the format of some entities in MovieLens does follow the normal linguistic rules, where their format is the concatenation of a few tokens instead of a group of separate tokens, which will bring a challenge when making the token-level context modeling for our C-BERT. That problem can be easily solved by further data processing to rectify the spelling. Even with this challenge, our proposed method still obtain a dramatic performance improvement in the ranking of the entity order. The correct entity is more likely to get a better ranking position with our proposed KIRS-CL.

CKE and CoFM show a performance drop compared to their basic KG component (i.e., TransR and TransE). One possible reason is that these models force their entity-aligned embeddings to satisfy the task of item recommendation. They need to find a trade-off between the recommendation task and the KG completion task. 

KTUP presents a good performance on KG completion by jointly training the recommendation task and the KG completion task. Because the recommendation task and the KG completion task are all based on TransH, it can be regarded as projecting all the embeddings into the same hyperplane. These two tasks can get more beneficial improvements under this setting, hence achieving satisfactory performance.

The large margin improvements of KIRS-CL on the KG completion task verifies its superior property in identifying the unseen link connection in KG and alleviating the influence of the KG incompleteness. The revised link connection could provide more information from KG to optimize the overall recommendation performance.

\section{Conclusion}
In this paper, we have constructed a new framework that unifies the semantic and structural representations of KG to understand user preferences better and generate accurate recommendations.
Regarding the semantic representation and the graph structure representation, we separately design a C-BERT and a CompGCN to get a better representation learning. To fully take advantage of the generated representation, we propose a warm-up training strategy with contrastive learning to make our system can be generalized to the cold-start application scenario. Extensive experiments on two real-world datasets demonstrate the superior performance of our proposed KIRS-CL in both warm- and cold-start recommendation scenarios. Also, a further link prediction experiment demonstrates the effectiveness of our proposed KIRS-CL in revising the missing link over KG, which explains the performance improvement on recommendation tasks.

\begin{acks}
This work is supported by the Australian Research Council under the streams of Future Fellowship (Grant No. FT210100624), Discovery Project (Grant No. DP190101985), Discovery Early Career Research Award (Grant No. DE200101465 and No. DE230101033), and Industrial Transformation Training Centre(No. IC200100022).
\end{acks}

\bibliographystyle{ACM-Reference-Format}
\bibliography{sample-base}


\end{document}